\documentstyle[12pt]{article}
\input epsfig.sty
\def\beq{\begin{equation}}
\def\eeq{\end{equation}}
\def\bea{\begin{eqnarray}}
\def\eea{\end{eqnarray}}

\def\NP{{\it Nucl. Phys.} }
\def\PL{{\it Phys. Lett.} }

\begin {document}
\begin{titlepage}
December 1996 \\
\begin{flushright}
HU Berlin-EP-96/62\\
\end{flushright}
\mbox{ }  \hfill hepth@xxx/9612120
\vspace{6ex}
\Large
\begin {center}
\bf{Nonabelian gauge field dynamics on matrix D-branes}
\end {center}
\large
\vspace{3ex}
\begin{center}
H. Dorn
\footnote{e-mail: dorn@qft3.physik.hu-berlin.de}
\end{center}
\normalsize
\it
\vspace{3ex}
\begin{center}
Humboldt--Universit\"at zu Berlin \\
Institut f\"ur Physik, Theorie der Elementarteilchen \\
Invalidenstra\ss e 110, D-10115 Berlin
\end{center}
\vspace{6 ex }
\rm
\begin{center}
{\bf Abstract}
\end{center}
We construct a calculational scheme for handling the matrix ordering
problems connected with the appearance of D-brane positions taking values
in the same Lie algebra as the nonabelian gauge field living on the 
D-brane. The formalism is based on the use of an one-dimensional
auxiliary field living on the boundary of the string world sheet and
taking care of the order of the matrix valued fields. The resulting system 
of equations of motion for both the gauge field and the D-brane position is 
derived in lowest order of the $\alpha ^{\prime}$-expansion.
\vspace{3ex}
\end{titlepage}
\setcounter{page}{1}
\pagestyle{plain}
\section{Introduction}
Dirichlet branes, i.e. hypersurfaces to which the endpoints of strings
are confined, play a fundamental role in the recent developments of the
string theory duality pattern \cite{pol, polwit, polrev} and the connected 
genesis of a unique underlying theory. Such D-branes arise from open 
strings with free ends by T-dualizing type I string theory 
or have to be added to type II theories in order to fulfill all the
duality requirements. In $\sigma $-model language the boundary of the
string world sheet couples to a gauge field $A_M(Y)$ living on the D-brane.
Equations of motion for this field, the brane position $f^{\mu}(Y)$ as well 
as the remaining target space fields can be obtained from the conformal
invariance condition of the 2D field theory. For abelian $A$ and $f$ this
program has been performed in ref. \cite{leigh}. The equations of motion
for these two fields turn out to be equivalent to the stationarity condition
of the Born-Infeld action
constructed out of the field strength related to $A$ and the metric and 
antisymmetric tensor on the brane induced from the target space.
\footnote{On some subtleties concerning this statement we will comment in
a forthcoming paper \cite{doproc}.}

For the generalization to the case of a nonabelian gauge symmetry it is
natural to assign values out of the same Lie algebra to both the gauge field 
and the brane position. This is due to the following property of the 
T-duality image of open strings with free ends coupled to an abelian gauge 
field. The gauge field components in the isometry directions become just
the fields describing the position of the D-brane in the dual theory.
An extensive discussion of this point is contained in \cite
{alv,do,munich,bolo}, examples for applications are found in \cite{ba}. 
The formal extension of this rule to nonabelian gauge
fields leads to the notion of a matrix valued brane and string endpoint
position \cite{wit, polrev}, respectively. If one tries to apply this c
oncept e.g. to the Yang-Mills equation on the D-brane
\footnote{We look at lowest order in $\alpha ^{\prime}$. In spite of some 
guesses \cite{callan3, nappi} there exists no proven nonabelian 
generalization of the Born-Infeld action \cite{tsbi}.}
\beq
h^{MN}~\hat{D}_M~F_{NL}~=~0~,
\label{a}
\eeq
one immediately runs into a serious ordering problem. $h^{MN}$ as the
inverse induced metric on the D-brane depends on the matrix valued
brane position and becomes a matrix with respect to the gauge indices.
The ordering problem concerns the construction of this matrix per se, 
as well as its ordering with respect to $\hat D F$.

The main objective of this paper is to decide this ordering problem
on the basis of a well defined calculational scheme and to demonstrate
the possibility of explicit calculations within this scheme. 

As a guiding principle for our formulation of the $\sigma $-model describing
strings coupled to matrix D-branes we require 
T-duality equivalence to a theory with open strings having free endpoints
for the special situation of target space fields independent of the 
coordinates orthogonal to the brane. The string couples in the bulk of its
world surface to the target space metric $G$, the dilaton $\Phi $ and an
antisymmetric tensor field $B$. After the formulation of the model we drop 
$\Phi $ in the following lowest order calculations, since at this level its 
effect is governed by classical considerations. Our model is designed to 
describe the NS-sector of the T-dual of type I strings $(B=0)$ or the
string D-brane interaction in type II theories $(B\neq 0)$. We will not 
discuss the extension to include RR bosonic fields along the line of 
\cite{schm}. 

To introduce notations and the use of the $\zeta $-auxiliary field
formalism we summarize in section 2 the result of the path integral
treatment of T-duality given in \cite{do}.
\footnote{There is a notational inconsistency in section 3 of the hard copy
of this paper, which has been corrected in the electronic version.} 
The next section is devoted to a 
definition  of the matrix D-brane model in a covariant manner and for
generic target space and brane fields. The following calculation of RG 
$\beta $-functions is done in analogy to \cite{leigh}. The main new aspects
concern the consequent bookkeeping of all the effects due to the presence
of the auxiliary field and the handling of explicit boundary parameter
dependent Dirichlet conditions in the intermediate steps of the calculation.
In this sense section 4 delivers the necessary formulae for the expansion of 
the action around a classical configuration obeying both the equations of
motion in the bulk as well as the boundary conditions. In section 5 we
discuss the propagator of the quantum field describing the fluctuations
of the string world sheet and the effects of the auxiliary field 
perturbation theory on the counter term evaluation. The conclusions will add 
some remarks concerning interpretation and work to be done.

\section{Functional integral derivation of the dual $\sigma $-model}
Our original $\sigma $-model describes an open string coupling in the bulk
to the target space metric $ G_{\mu \nu}$, an antisymmetric tensor
$B_{\mu \nu}$ and the dilaton $\Phi $ (collective notation by $\Psi =
(G,B,\Phi)$). In
addition it couples via its ends to a nonabelian gauge field $A_{\mu}$
taking its values in the Lie algebra of a nonabelian gauge group ${\cal G}$.
We assume the existence of one Killing vector $k^{\mu}(X)$ and the invariance
of our model under the corresponding diffeomorphism. Choosing adapted
coordinates all target space fields are independent of $X^0$
(For $A$ a gauge transformation may be necessary to reach this
conclusion \cite{do}.)
\beq
X^{\mu}=(X^0,X^M),~~~k^{\mu}=(1,0),~~~\partial _0 \Psi=0,~~~\partial _0 A_
{\mu}=0~.
\label{1}
\eeq
The partition function is given by
\beq
Z[\Psi ,A]~=~\int DX^{\mu}e^{iS[\Psi ,C=0;X]}~\mbox{tr}Pe^{i\int _
{\partial M}A_{\mu}dX^{\mu}}~.
\label{2}
\eeq
To streamline notation we have expressed $Z$ in terms of an action $S$
which below is allowed to depend on an abelian vector field
$C_{\mu}(X^M,s)$ with possibly explicit dependence on the parameter
on $\partial M$
\bea
S_M[\Psi ;X]&=&\frac{1}{4\pi \alpha ^{\prime}}\int _M d^2 z \sqrt{-g}
\left ( \partial _m X^{\mu} \partial _n X^{\nu} E^{mn}_{\mu \nu}(X(z))+
\alpha ^{\prime} R^{(2)}\Phi (X(z)) \right )~,\nonumber\\
S[\Psi ,C;X]&=&S_M[\Psi ;X]~+~\int_{\partial M} \left (C_{\mu}(X^M(z(s)),s)\dot X^{\mu}-\frac{1}{2\pi}
k(s)\Phi \right )ds~,\nonumber \\
E^{mn}_{\mu \nu}(X)&=&g^{mn}G_{\mu \nu}(X)~+~\frac{\epsilon ^{mn}}{\sqrt{-g}}
B_{\mu \nu}(X)~.
\label{3}
\eea
$R^{(2)}$ is the curvature scalar on the 2D manifold $M$, $k(s)$ the
geodesic curvature on its boundary $\partial M$ parametrized by $z(s)$.

To disentangle the path ordering implied by the Wilson loop we introduce an
one-dimensional auxiliary field
$\zeta _a (s)$ living on the boundary $\partial M$ \cite{zform, fort}. It
has the propagator
\beq
\langle \bar{\zeta}_a (s_1)\zeta _b (s_2)\rangle _0=\delta _{ab}
\Theta (s_2-s_1)
\label{4}
\eeq
and  couples to $X^{\mu}$ via the interaction term
\beq
i\bar{\zeta}_a A_{\mu}^{ab}(X(z(s)))\zeta _b (s)
\partial _m X^{\mu}\dot{z}^m(s)~.
\label{5}
\eeq
Then we can write (choosing $0\leq s\leq 1$)
\beq
Z=\int DX^{\mu}~D\bar{\zeta}~D\zeta ~\bar{\zeta}_a(0)\zeta _a(1)
e^{iS_0[\bar{\zeta}, \zeta ]}\exp (iS[\Psi ,\bar{\zeta}A
\zeta ; X]).
\label{6}
\eeq

Under the $\zeta $-path integral we can repeat the dualization procedure
for abelian $A$ \cite{do}. Due to the presence of
$\zeta (s)$ in $C_{0}(X^M(z(s)),s)= \bar \zeta (s)A_0(X^M(z(s)))\zeta (s)$ the
resulting Dirichlet condition depends on $s$ explicitly. With the help of the
functional
\beq
{\cal F}[\Psi,C\vert f]~=~\int _{X^0(z(s))=f(X^M(z(s)),s)} D X^{\mu}~\exp (iS
[\Psi ,C;X])
\label{7}
\eeq
we can write the result for $Z$ as
\beq
Z~=~ \int D\bar{\zeta}~ D\zeta ~\bar{\zeta}_b (0)\zeta _b (1)e^{iS_0 [
\bar{\zeta},\zeta ]}~{\cal F}[\tilde{\Psi},\bar{\zeta}
\tilde A \zeta \vert -2\pi\alpha ^{\prime}\bar{\zeta}A_0
\zeta ]~.
\label{8}
\eeq
The dual target space fields are given by
\beq
\tilde A_{\mu}~=~(0,A_M)
\label{9}
\eeq
and the standard Buscher rules \cite{buscher} for $\tilde{\Psi}
\leftrightarrow \Psi$.\\ 

For abelian $A$ the boundary condition for the dual model (note $X^M=\tilde
X^M$) is \cite{do}
$$\tilde X^0(z(s))=-2\pi\alpha ^{\prime} A_0(\tilde X^M(z(s)))$$
which constrains the end points of the string to the hypersurface
$\tilde X^0=-2\pi\alpha ^{\prime} A_0(\tilde X^M)$ with free movement inside
this hypersurface (D-brane). In contrast a general $s$-dependent boundary
condition as in (\ref{7}) has no D-brane interpretation. However, we have
to keep in mind that in (\ref{8}) we need only boundary conditions
of the type 
$$\tilde X^0(z(s))=-2\pi\alpha ^{\prime}\bar{\zeta}(s) A_0
(\tilde X^M(z(s)))\zeta (s)~.$$ 
This type still allows a D-brane
interpretation: The brane as a whole changes its position in the target space
in dependence on $s$. 

The $\zeta $-integration results in ordering the matrices sandwiched between
$\bar{\zeta}(s)$ and $\zeta (s)$ with respect to increasing $s$. But after
performing the functional integral over the world surfaces $X^{\mu}(z)$
there is no longer any correlation between a given target space point and
$s$. The situation is improved if one treats (\ref{7}) and (\ref{8})
within the background field method ($bm$). Then both in ${\cal F}_{bm}$
and $Z_{bm}$ all dependence on target space coordinates is realized via
a classical string world sheet configuration $X_{cl}$. The result of the 
$\zeta $-integration is then \cite{do}
\beq
Z_{bm}[\Psi ,A;X_{cl}]=\mbox{tr}P{\cal F}_{bm} [\tilde{\Psi}, \tilde{A};
\tilde X_{cl}\vert -2\pi \alpha ^{\prime}A_0]~.
\label{11}
\eeq
Path ordering now refers to the classical path $\tilde X_{cl}(z(s))$ and
involves both the matrices appearing in the second argument
of ${\cal F}_{bm}$ as well as $A_0$ entering via the argument specifying the
boundary condition.

The insertion of a matrix as boundary condition is performed $after$
${\cal F}_{bm}$ has been calculated with scalar (not matrix valued) 
$s$-dependent boundary condition. In this formalism we can avoid wondering 
about the target space interpretation of matrix valued boundaries. The 
situation is similar to dimensional regularization, the change from integer 
$n$ to complex $n$ is performed $after$ $\int d^n x ...$ has been calculated
for integer $n$.
\section{Covariant definition of the matrix D-brane model}
In analogy to \cite{leigh} we now generalize our consideration to the
case of several abelian isometries, define the resulting dual model
in a covariant way and extend it to generic target space fields $\Psi $
at the end. For this purpose we describe a matrix D-brane with a
p-dimensional world hypersurface by matrix valued target space coordinates
$$f^{\mu}(Y ^N)~,~~~~N=1,...,p~, $$
taking their values in the Lie algebra of ${\cal G}$.
The open string under discussion couples in the bulk as usual to the target
space fields $\Psi $. In addition there is a nonabelian gauge field
$A_M(Y)$ living on the D-brane \footnote{This field corresponds to
$\tilde A_N(X^N)=A_N(X^N)$ in the previous section.}.
Under a gauge transformation with $\Omega (Y)\in{\cal G}$ the field
$A$ transforms as a standard gauge field while $f$ transforms homogeneously
$f^{\mu}\rightarrow \Omega f^{\mu}\Omega ^{-1}$.
The Dirichlet boundary condition as well as the coupling of the string world
sheet boundary to $A_M$ will be formulated with the help of the
one-dimensional auxiliary field $\zeta$ of the previous section.

Let again $z(s)$ parametrize the string world sheet boundary in 2D parameter
space. The Dirichlet boundary condition relates the target space image
$X^{\mu}(z(s))$ of this $z$-space curve to a curve on the D-brane
$Y^N (s)$
\beq
X^{\mu}(z(s))~=~ \bar{\zeta}(s)f^{\mu}(Y^N (s))\zeta (s)~.
\label{12}
\eeq
The relevant action is ($S_M$ from (\ref{3}))
\bea
{\cal S}[\Psi,A;\bar{\zeta},\zeta ;X]&=&S_M[\Psi ;X]~+~{\cal S}_{\partial M}~,
\nonumber\\
{\cal S}_{\partial M}&=&\int _{\partial M}\left
(\bar{\zeta}(s)A_N(Y(s))\zeta (s)\dot{Y}^N-
\frac{1}{2\pi}k(s)\Phi (X(z(s)))\right )ds~.
\label{13}
\eea
With the covariant version of (\ref{7})\footnote{We avoid to introduce a new
symbol for it.}
\beq
{\cal F}[\Psi,\bar{\zeta}A\zeta \vert \bar{\zeta}f\zeta ]~=~
\int _{(\ref{12})}DX^{\mu}\exp (i{\cal S}[\Psi,A;\bar{\zeta},\zeta ;X])
\label{14}
\eeq
the partition function is finally given by
\bea
{\cal Z}[\Psi ,A]&=& \int D\bar{\zeta}(s) D\zeta (s) ~\bar{\zeta}_b (0)\zeta _
b (1)e^{iS_0 [\bar{\zeta},\zeta ]}~{\cal F}[\Psi,\bar{\zeta}A \zeta
\vert \bar{\zeta}f\zeta ]\nonumber\\[2mm]
&=&\mbox{tr}P{\cal F}[\Psi ,A\vert f]~.
\label{15}
\eea

Before turning in the next sections to the calculation of lowest order RG
$\beta$-functions for the model defined above, we still provide the
background expansion of the boundary condition (\ref{12}).
If $X$ and $Y$ are varied around a configuration satisfying (\ref{12})
at fixed $\zeta ,~\bar{\zeta}$ we get a gauge non-covariant result.
Therefore, we combine the variation of $Y$ with an adapted gauge
transformation of $\zeta ,~\bar{\zeta}$ (${\cal C}$ denotes the
straight line connecting $Y$ and $Y+\delta Y$)
\bea
\zeta & \rightarrow & \zeta + \delta \zeta = P\exp (i\int _{\cal C}A_NdY^N)~
\zeta \nonumber\\
&&=\zeta + i\delta Y^M A_M \zeta - \frac{1}{2}\delta Y^M\delta Y^N
(A_MA_N-i\partial _M A_N)\zeta + O((\delta Y)^3)~,
\label{16}
\eea
which leads to
\beq
\delta X^{\mu}~=~\delta Y^M\bar{\zeta}D_Mf^{\mu}\zeta +\frac{1}{2}\delta
Y^M\delta Y^N \bar{\zeta}D_MD_Nf^{\mu}\zeta +O((\delta Y)^3)~.
\label{17}
\eeq
$D_M=\partial _M -i[A_M,\cdot ~] $ is the gauge covariant derivative.

Target space distances along curves subject to (\ref{12}),(\ref{16}) are
measured with the induced metric
\beq
h_{MN}(Y(s),s)~=~f^{\mu}_{;M}(Y(s),s)\cdot f^{\nu}_{;N}\cdot G_{\mu \nu}
(\bar{\zeta}(s)f(Y(s))\zeta (s))~,
\label{18}
\eeq
where we defined
\beq
f^{\mu}_{;M}(Y(s),s)~=~\bar{\zeta}(s)~D_Mf^{\mu}(Y(s))~\zeta (s)~.
\label{18a}
\eeq
In these formulae we have indicated the functional dependences in full detail,
since it will be important for later use to know at which places matrices
appear after performing the $\zeta$-integration.
Now we want to develop a calculus of covariant derivation with respect to this
induced metric, which is covariant with respect to the gauge group
${\cal G} $, too. Motivated by (\ref{16}) one has to replace the derivative
$\partial _N$ by
$$\partial _N -i\bar{\zeta}A_N\frac{\partial}{\partial \bar{\zeta}}
+i\zeta _bA^{ab}_N\frac{\partial}{\partial \zeta _a}~.$$
As far as $Y$-dependence appears via quantities sandwiched between $\bar
{\zeta}$ and $\zeta $, the final recipe is to replace $\partial _N$ by
$D_N$ applied to the quantity under consideration sandwiched between
$\bar{\zeta}~,\zeta$. Then the Levi-Civita connection related to (\ref{18})
is
\bea
\gamma _{MNL}&=&\frac{1}{2}[(MN,L)+(ML,N)+(NM,L)+(NL,M)-
(LM,N)-(LN,M)]\nonumber\\
&+&f^{\mu}_{;M}\cdot f^{\nu}_{;N}\cdot f^{\lambda}_{;L}\cdot
\Gamma _{\mu\nu\lambda}(\bar{\zeta}f\zeta )~,
\nonumber\\[2mm]
(MN,L)&=&\bar{\zeta}(s)D_MD_Nf^{\nu}(Y(s))\zeta (s)\cdot f^
{\lambda}_{;L}\cdot G_{\nu\lambda}(\bar{\zeta}f\zeta )~,
\label{18b}
\eea
with $\Gamma_{\mu\nu\lambda}$ denoting the target space connection
coefficients.

Introducing Riemann normal coordinates $\xi ^{\mu} $ and $\eta ^N$
for the target space and the brane, respectively, we get finally
\beq
\xi ^{\mu}~=~\eta ^M~f^{\mu}_{;M}~+~\frac{1}{2}~\eta ^M\eta ^N
K^{\mu}_{MN}~+~O(\eta ^3)~,
\label{19}
\eeq
with
\beq
K^{\mu}_{MN}~=~\bar{\zeta}D_MD_Nf^{\mu}\zeta~+~f^{\alpha}_{;M}
\cdot f^{\beta}_{;N}\cdot\Gamma ^{\mu}_{\alpha \beta}(\bar{\zeta}f
\zeta )~-~\gamma ^A_{MN}~f^{\mu}_{;A}~.
\label{20}
\eeq
\section{Background field expansion of the action}
We expand around a classical string configuration satisfying the stationarity
condition both in the bulk of the string world surface $M$ (equation of
motion) and on the boundary $\partial M$ (generalized Neumann boundary
condition).
\footnote{It is a consequence of free movement of the string endpoints
inside the D-brane and is compatible with the Dirichlet condition (\ref{12})
which forbids movement orthogonal to the D-brane \cite{leigh}.} The equation
of motion for $X$ (We drop the index $cl$ used in section 2.) is taken into 
account by dropping linear terms in $\xi$ in the bulk. The generalized Neumann
boundary condition has to be written down explicitly, since it will need
some discussion below (${\bf t}$ and ${\bf n}$ denote tangential and normal
components, respectively.) \footnote{Since we are performing lowest order
calculations only, we will consider flat 2D metric and drop the dilaton 
field $\Phi $ in the following.}
\beq
\partial _{\bf n}X^{\alpha}G_{\alpha\mu}\bar{\zeta}D_Mf^{\mu}\zeta
-\partial _{\bf t}X^{\alpha}B_{\alpha\mu}\bar{\zeta}D_Mf^{\mu}\zeta
-2\pi \alpha ^{\prime}\partial _{\bf t}Y^A\bar{\zeta}F_{AM}\zeta~=~0~.
\label{21}
\eeq
Now the standard expansion of $S_M[\Psi ;X+\delta X]$ gives (e.g.\cite{curt, do86, leigh})
\bea
S_M&=&S_M[\Psi ;X]+\frac{1}{4\pi \alpha ^{\prime}}\int _M d^2z~(G_{\alpha
\beta}\nabla _m\xi ^{\alpha}\nabla ^m\xi ^{\beta}+\xi ^{\alpha}\xi ^{\beta}
R_{\mu\alpha\beta\nu}\partial _mX^{\mu}\partial ^mX^{\nu}\nonumber\\
&&~~~~~~~~~~~~-\frac{1}{2}\xi ^{\alpha}\xi ^{\beta}\nabla _{\alpha}
H_{\mu\beta\nu}
\epsilon^{mn}\partial _mX^{\mu}\partial _nX^{\nu}+\xi ^{\alpha}\nabla _m \xi^
{\beta}H_{\alpha\beta\mu}\partial _nX^{\mu}\epsilon ^{mn} )\nonumber\\
&+&\frac{1}{2\pi\alpha ^{\prime}}\int _{\partial M} (\partial _{\bf n}
X^{\alpha}G_{\alpha\beta}\xi^{\beta}-\partial _{\bf t} X^{\alpha}
B_{\alpha \beta}\xi^{\beta}\nonumber\\
&&~~~~~~~~+\frac{1}{2}\xi ^{\alpha}\xi ^{\beta}\nabla _{\alpha}B_{\beta\mu}
\partial _{\bf t}X^{\mu}+\frac{1}{2}\xi ^{\alpha}\nabla _{\bf t}\xi ^{\beta}
B_{\alpha\beta})ds~+~O(\xi ^3)~.
\label{22}
\eea
$\nabla $ denotes the target space covariant derivative. In the expansion of
the the gauge field dependent part ${\cal S}_{\partial M}$
in (\ref{13}) use has to be made of the $\bar{\zeta},~\zeta $ quantum
equation of motion including all contact terms. To shorten the treatment
we use instead the standard variation formulae for the Wilson loop
(see e.g. \cite{fort} and refs. therein) and re-express them afterwards
in the auxiliary field language. For
$$\mbox{tr}PU[Y]~=~\mbox{tr}P\exp (i\int _{\partial M}A_MdY^M)$$
one has up to order $O((\delta Y)^3)$
\bea
\mbox{tr}PU[Y+\delta Y]&=&\mbox{tr}P\left ( U[Y]\exp \left (i\int ds(F_{MN}
\dot{Y}^N\delta Y^M+\frac{1}{2}D_MF_{NK}\dot{Y}^K\delta Y^M\delta Y^N
\right .\right .
\nonumber\\
&&~~~~~~~~~~~~~~~~~~~~~
\left .\left .+\frac{1}{2} F_{MN}\delta Y^M\delta \dot{Y}^N\right )
\right )~.
\label{23}
\eea
The translation into $\zeta$-language is
\bea
\lefteqn{\mbox{tr}PU[Y+\delta Y]
~=~\int D\bar{\zeta}D\zeta ~\bar{\zeta}_b(0)\zeta _b(1) e^{iS_0+i{\cal S}
_{\partial M}[Y]}}\label{24}\\
&&\cdot ~\exp \left (i\int ds~\bar{\zeta}(F_{MN}\dot{Y}^N
\delta Y^M+\frac{1}{2}D_MF_{NK}\dot{Y}^K\delta Y^M\delta Y^N
+\frac{1}{2} F_{MN}\delta Y^M\delta \dot{Y}^N)\zeta \right )~.
\nonumber
\eea
On the other side by comparing
$$\mbox{tr}PU[Y+\delta Y]=\int D\bar{\zeta}D\zeta ~\bar{\zeta}_b(0)
\zeta _b(1)e^{iS_0+i{\cal S} _{\partial M}[Y+\delta Y]}$$
with (\ref{24}) we get the wanted expansion for ${\cal S}_{\partial M}$.
In writing its final version we still express $\delta Y$ in terms of the
Riemann normal coordinates on the brane $\eta $ and denote the covariant
derivative with respect to both the gauge group ${\cal G}$ and the
induced metric by $\hat D _M$
\bea
{\cal S}_{\partial M}[A;Y+\delta Y]&=&{\cal S}_{\partial M}[A;Y]
\label{25}\\
&+&\int ds~\bar
{\zeta}(F_{MN}\dot{Y}^N
\eta ^M+\frac{1}{2}\hat D_MF_{NK}\dot{Y}^K\eta ^M\eta ^N
+\frac{1}{2} F_{MN}\eta ^M\hat D_{\bf t} \eta ^N)\zeta~.
\nonumber
\eea
The sum of (\ref{22}) and (\ref{25}) yields the expansion of ${\cal S}$.
For further simplification we imply the boundary condition (\ref{21}) for the
classical background configuration $X,~Y$, eliminate on $\partial M$ $\xi $
by (\ref{19}) in favour of $\eta $ and introduce in analogy to (\ref{18})
\beq
b_{MN}(Y(s),s)~=~f^{\mu}_{;M}(Y(s),s)\cdot f^{\nu}_{;N}\cdot B_{\mu \nu}
(\bar{\zeta}(s)f(Y(s))\zeta (s))~.
\label{26}
\eeq
Then, using
\bea
\partial _{\bf t}X^{\alpha}&=&~\dot Y^A~f^{\alpha}_{;A}\nonumber\\
\nabla _{\bf t}\xi ^{\beta}&=&\hat D_{\bf t}\eta ^M~f^{\beta}_{;M}
+\eta ^M\dot Y^A~\bar{\zeta}\hat D_AD_Mf^{\beta}\zeta +\eta ^M\dot Y^A
\Gamma ^{\beta}_{\alpha \mu}~f^{\alpha}_{;A}~f^{\mu}_{;M}
\label{27}
\eea
and
\beq
\hat D_M~b_{NA}~=~\bar{\zeta}\hat D_MD_Nf^{\nu}\zeta ~f^{\alpha}_{;A}~B_
{\nu\alpha}+f^{\nu}_{;N}~\bar{\zeta}\hat D_MD_Af^{\alpha}\zeta ~B_
{\nu\alpha}+f^{\nu}_{;N}~f^{\alpha}_{;A}~f^{\mu}_{;M}~\partial _{\mu}
B_{\nu\alpha}~,
\label{28}
\eeq
we arrive up to $O(\xi ^3,\eta ^3)$ at
\bea
{\cal S}&=&{\cal S}[\Psi,A;\bar{\zeta},
\zeta ;X]\nonumber\\
&+&\frac{1}{4\pi \alpha ^{\prime}}\int _M d^2z~\left (G_{\alpha
\beta}\nabla _m\xi ^{\alpha}\nabla ^m\xi ^{\beta}+\xi ^{\alpha}\xi ^{\beta}
R_{\mu\alpha\beta\nu}\partial _mX^{\mu}\partial ^mX^{\nu}\right .\label{29}\\
&&~~~~~~~~~~~-\left .\frac{1}{2}\xi ^{\alpha}\xi ^{\beta}\nabla _{\alpha}
H_{\mu\beta\nu}\epsilon^{mn}\partial _mX^{\mu}\partial _nX^{\nu}+\xi ^
{\alpha}\nabla _m \xi^
{\beta}H_{\alpha\beta\mu}\partial _nX^{\mu}\epsilon ^{mn}\right )\nonumber\\
&+&\frac{1}{4\pi \alpha ^{\prime}}\int _{\partial M}\left (\eta ^M\eta ^N
\partial _{\bf n}X^{\alpha}G_{\alpha\beta}K^{\beta}_{MN}+\eta ^M\eta ^N\dot
Y^A(\hat D_Mb_{NA}+2\pi\alpha ^{\prime}\bar{\zeta}\hat D_MF_{NA}\zeta)
\right .\nonumber\\&&
~~~~~~-\left .i\eta ^M\eta ^N\dot Y^A~\bar{\zeta}[F_{AM},f^{\alpha}]
\zeta ~f^{\nu}_{;N}~B_{\nu\alpha}
+\eta ^M\hat D_{\bf t}\eta ^N(b_{MN}+2\pi\alpha ^{\prime}\bar{\zeta}F_{MN}
\zeta )\right )ds~.
\nonumber
\eea
Comparing this result with the abelian case \cite{leigh} we find just
one additional structure, the $[F,f]$ commutator term. All other modifications
refer only to the appearance of the auxiliary fields. Note that in (\ref{29})
besides the explicit $\zeta$'s there is more $\zeta $-dependence
in $G,~B$, since the arguments of these fields on $\partial M$ are given by
(\ref{12}), and in $b,~\hat D,~f^{\nu}_{;N}$ and $K$ via 
(\ref{26},\ref{18b},\ref{18a},\ref{20}).
\section{Lowest order calculation of RG $\beta $-functions}
To begin with we need the propagator for the quantum corrections $\xi $
and their manifestation on the brane $\eta $. The relation between $\xi $ and
$\eta $ is a consequence of the Dirichlet condition implied as an external
constraint on the integrand of the functional integral in (\ref{14}). The
choice of any further boundary condition, to make the propagator $\langle
\xi ~\xi \rangle _0 $
well defined, is a matter of technical convenience only. It has implications
on the question concerning the set of boundary vertices contributing in the
perturbative evaluation. For instance, the use of a propagator obeying
the Neumann condition including the Lorentz force has allowed in
\cite{callan1, callan2} to sum in one graph all orders of $\alpha ^{\prime}$
for the case of constant background fields. Unfortunately, we cannot repeat
this trick for our case since the price we paid for handling the nonabelian
structures is the explicit boundary parameter dependence. We did not succeed
in constructing the corresponding propagator explicitly. However, to
calculate counter-terms in lowest order it is necessary to know the short
distance behaviour only:
\bea
\langle\xi ^{\mu}(z_1)\xi^{\nu}(z_2)\rangle _0&=&D^{\mu\nu}(z_1,z_2)~+~N^
{\mu\nu}(z_1,z_2)\nonumber\\
&=&-\alpha ^{\prime}\left (G^{\mu\nu}(X(z_2))+O(z_1-z_2)
\right )\cdot\log \vert z_1-z_2\vert ~.
\label{30}
\eea
The last line is valid inside $M$. The boundary behaviour is controlled by
\bea
D^{\mu\nu}(z_1,z_2)&=&0,~~~~\mbox{if}~z_1~\mbox{or}~z_2 \in \partial M~,
\nonumber\\
\nabla _{\bf n} N^{\mu\nu}&=&0~~~~\mbox{on}~~ \partial M~, \nonumber\\
N^{\mu\nu}(z(s_1),z(s_2))&=&-2\alpha ^{\prime}\left (f^{\mu}_{;M}h^{MN}
f^{\nu}_{;N}+O(s_1-s_2)\right )\cdot\log\vert s_1-s_2\vert ~.
\label{31}
\eea
The arguments of $f^{\mu}_{;M}$ and $h^{MN}$ are $Y(s_2),s_2$. Eqs. (\ref{31})
and (\ref{19}) also imply\footnote{The factor 2 of the boundary singularity
relative to the bulk singularity has been discussed at length in \cite{bd}.}
\beq
\langle\eta ^A(s_1)\eta ^{B}(s_2)\rangle _0~=~-2\alpha ^{\prime}\left (h^{AB}+
O(s_1-s_2)\right )\cdot\log\vert s_1-s_2\vert~.
\label{32}
\eeq
The existence of such a propagator is guaranteed at least in the vicinity
of constant background fields and no explicit $s$-dependence \cite{leigh}.
We assume that there are no global obstructions. The simple Neumann condition
guarantees that the boundary vertex $\xi\nabla _{\bf n}\xi G$
arising from the partial integration of the kinetic term in (\ref{29}) does
not contribute.

The bookkeeping of divergent diagrams in the perturbative evaluation of
the partition function ${\cal Z}$ is the same as in the case of a string with
free ends \cite{curt,do86,callan1,callan2}. There are only modifications
in the classical factors multiplying the quantum fields in the vertices.
In this paper we are interested in counter terms located on the boundary
$\partial M$, exclusively. Then at one-loop level we have to consider the
tadpole graph constructed with the $\eta\eta$-vertex (corresponding to the
first three terms in the boundary integral of (\ref{29})) and the
bulk-boundary interference graph \cite{do86,callan2,leigh} constructed
with the $\eta
\hat D_{\bf t}\eta $-vertex (last term in the boundary integral) and
the $\xi \nabla \xi$-vertex (last term in the bulk integral). In
addition all diagrams constructed out of these two basic diagrams by
multiple insertion of the  $\eta\hat D_{\bf t}\eta $-vertex contribute.
However, since the propagator is not known explicitly, we skip the otherwise
possible summation \cite{callan1,leigh} and restrict ourselves to the
above mentioned two basic diagrams. Altogether, then our final result
will represent the beginning of an expansion in $\alpha ^{\prime}$ and
$B_{\mu\nu}$.

In this sense the tadpole contribution to the counter-term action is
($\Lambda $ denotes the short distance cutoff.)
\bea
\frac{1}{2\pi}\log\Lambda \cdot h^{MN}\left (K^{\alpha}_{MN}G_{\alpha
\beta}\partial _{\bf n}X^{\beta}\right .&+&\dot Y^A(\hat D_Mb_{NA}+2\pi\alpha ^
{\prime}\bar{\zeta}\hat D_MF_{NA}\zeta )\nonumber\\
&-&i\left .\dot Y^A~\bar{\zeta}[F_{AM},f^{\alpha}]\zeta ~f^{\nu}_{;N}~B_{\nu\alpha}\right )~.\nonumber
\eea
The bulk-boundary interference diagram yields
$$ -\frac{1}{2\pi}\log\Lambda \cdot \frac{1}{2}(b^{MN}+2\pi\alpha ^{\prime}
\bar{\zeta}F^{MN}\zeta )f^{\mu}_{;M}f^{\nu}_{;N}H_{\mu\nu\alpha}\partial
_{\bf n}X^{\alpha}~.$$
In total this gives for the boundary part of the counter-term action
up to higher orders in $\alpha ^{\prime}$ and $B$
\bea
\Delta {\cal S}&=&\frac{\log \Lambda}{2\pi}\int _{\partial M}\left (
(h^{MN}K_{MN}^{\mu}G_{\mu\nu}\right .+\frac{1}{2}(b^{MN}+2\pi\alpha
^{\prime}\bar{\zeta}F^{MN}\zeta )f^{\alpha}_{;M}f^{\beta}_{;N}H_{\alpha\beta
\nu} )\cdot\partial _{\bf n}X^{\nu}\label{33}\\
&&~~~~~~~~~~~+\left .h^{MN}(\hat D_Mb_{NA}+2\pi\alpha ^{\prime}
\bar{\zeta}\hat D_MF_{NA}\zeta -i\bar{\zeta}[F_{AM},f^{\alpha}]
\zeta ~f^{\nu}_{;N}~B_{\nu\alpha})\cdot\dot Y^A\right )ds~.\nonumber
\eea

Now the auxiliary field formalism has to do its last job. $\bar{\zeta}$ and
$\zeta $ appear in (\ref{33}) at many places, compare the last remark in
section 4. As a general rule they always sandwich some matrix. Since $\Delta
{\cal S}$ is local in $s$ we need a procedure to perform $\zeta $-integrals
of the type \footnote{Thinking e.g. in terms of Taylor expanded background
fields.}
\beq
I_{ab}~=~\int D\bar{\zeta}D\zeta~ \exp \left (iS_0+i\int\bar{\zeta}A_N
\zeta dY^N\right )~\bar{\zeta}_a(0)\zeta _b(1)~\prod _{j=1}^N\bar{\zeta}
{\cal M}_j\zeta (s)~.
\label{34}
\eeq
The generalization to the necessary multiple insertions of such vertices
is straightforward.

Due to the $\Theta $-function in the $\zeta $-propagator there are no
$\zeta $-loops going forward and backward in one-dimensional $s$-space.
Only $\zeta $-loops consisting out of propagators at coinciding points
i.e. $\zeta $-tadpoles are allowed. However, due to the ambiguity of
$\Theta (0)$ they are subject to renormalization ambiguities. If the
Wilson loop under the functional integral in (\ref{2}) is replaced
by the full two point function $\langle \bar{\zeta}_b(0)\zeta _b(1)\rangle $
(compare (\ref{6}))  we decide to specify the definition of the $\zeta $
quantum theory by restriction to the sector of diagrams in which the gauge
index flux is completely described by only one continuous line connecting $s=0$
and $s=1$. 
\begin{figure}[t]
\begin{center}
\mbox{\epsfig{file=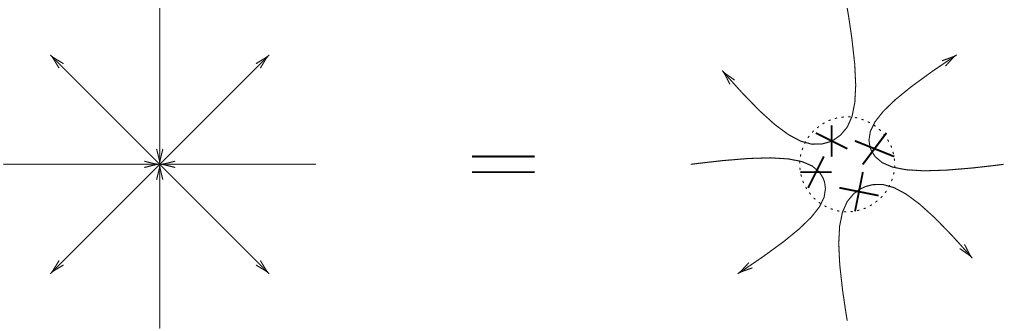, width=80mm}}
\end{center}
\noindent {\bf Fig.1}\ \ {\it
Gauge index ``fine structure'' of a local vertex of the type indicated
in (\ref{34})\\ 
\hspace*{12mm}for $N=4$. The crosses denote a matrix.}
\end{figure}
This concept is illustrated by fig.1 and fig.2, it is based on the fact
that all vertices appearing in the calculations have a gauge index structure
as in (\ref{34}). 
\begin{figure}[b]
\begin{center}
\mbox{\epsfig{file=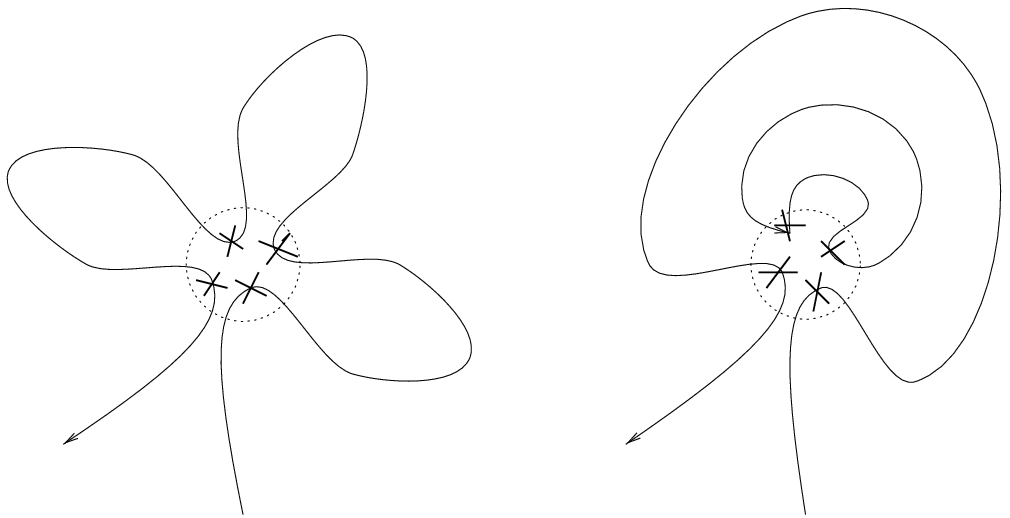, width=80mm}}
\end{center}
\noindent {\bf Fig.2}\ \ {\it Allowed and forbidden contributions. Use is
made of the vertex notation\\
\hspace*{12mm}introduced in fig.1}
\end{figure}
Via inserted matrices it connects legs in a pairwise
manner. Just this restriction we imply on the covariant definition
of our model in section 3, too.

After this specification we see that in the evaluation of $I$ the sum of all
permuted products of the matrices ${\cal M}_j$ appears
$$I_{ab}~=~N!~(\Theta (0))^{N-1}~ \left (P(U~\mbox{Sym}(\prod _{j=1}^{N}
{\cal M}_j))\right )_{ab}~.$$
At this point we still have a renormalization ambiguity. We fix the ambiguity
in $\Theta (0)$ by requiring that the whole formalism reproduces in the
abelian case the then well known results. This means to fix depending 
the number of factors $N$: $(\Theta (0))^{N-1}=(N!)^{-1}$.
Altogether we found for our local vertices the identification
\beq
\prod _{j=1}^N\left (\bar{\zeta}{\cal M}_j\zeta \right )\simeq
\bar{\zeta}~\mbox{Sym}\left (\prod _{j=1}^N {\cal M}_j \right )~\zeta ~.
\label{35}
\eeq
We define an operation ${\cal Q}$ by
\beq
{\cal Q}\{\prod _{j=1}^N\left (\bar{\zeta}{\cal M}_j\zeta \right )\}
~=~\mbox{Sym}\left (\prod _{j=1}^N {\cal M}_j \right )~.
\label{36}
\eeq
and extend it linearly to sums of products of sandwiched matrices.

The RG $\beta $-functions for the gauge field $A$ and the brane position $f$
can be read off from (\ref{33}). Using the just defined operation ${\cal Q}$
the boundary part of the conformal invariance condition then becomes
\footnote{We assume that as usual in lowest order there is no difference
between the RG $\beta $-functions and the Weyl anomaly coefficients
\cite{tsweyl}.}
\bea
{\cal Q}\{ h^{MN}(\hat D_Mb_{NA}+2\pi\alpha ^{\prime}\bar{\zeta}\hat D_MF_{NA}
\zeta -i\bar{\zeta}[F_{AM},f^{\alpha}]\zeta ~f^{\nu}_{;N}~B_{\nu\alpha})
\}&=&0\nonumber\\
{\cal Q}\{ h^{MN}K_{MN}^{\mu}G_{\mu\nu}+\frac{1}{2}(b^{MN}+2\pi\alpha
^{\prime}\bar{\zeta}F^{MN}\zeta )f^{\alpha}_{;M}f^{\beta}_{;N}H_{\alpha\beta
\nu} \}&=&0 ~.\label{37}
\eea
After the application of ${\cal Q}$ there is present no longer any
$\bar{\zeta}$ or $\zeta $. The manner how matrices were sandwiched between
the auxiliary fields determines which matrices have to be handled as basic
entities under the symmetrization procedure. E.g. $\hat D_MF_{NA}$ and
$[F_{AN},f^{\alpha}]$ appear as such basic matrices. Therefore, the
commutators due to the nonabelian gauge structure will not be removed by the
symmetrization.

We interpret our main result (\ref{37}) as the system of equations of
motion for the gauge field living on the brane $and$ the brane position.
At this stage we leave open the question whether e.g. the first equation
is $the$ gauge field equation or whether as in \cite{leigh} a certain
linear combination of the two equations plays this role. Such a linear
combination arises if one uses the boundary condition (\ref{21}) to
transform the projection onto the brane of the $\partial _{\bf n}X^{\nu}$ 
term in (\ref{33}) into a $\dot Y^A$ term. For more discussion
on this point we refer to a paper in preparation \cite{doproc}.

The structure $[F,f]B$ is a genuine nonabelian effect. It leads for $B\neq 0$
to a direct interaction between the gauge field and the brane position.
For gauge groups whose Lie algebras contain multiples of the identity the
whole system is translation invariant in target space, otherwise this
invariance is broken.
\section{Conclusions}
The use of the one-dimensional auxiliary field formalism allowed us to 
express the partition function for an open string with free ends coupling
to a nonabelian gauge field after a T-duality transformation as certain 
functional integral over the auxiliary field (\ref{9}). The integrand
contains the functional ${\cal F}$ which corresponds to the partition function
of a theory which obeys explicit boundary parameter dependent Dirichlet
conditions. As discussed at the end of section 2, the replacement of the
function specifying the boundary condition by a matrix valued object 
(enforced by the auxiliary field integration) gives meaning to
the notion of matrix valued D-brane position. Motivated by this consequence
of T-duality we gave, again using the auxiliary field, a definition of
a model describing a string bound with its ends to a matrix D-brane
and coupling to a nonabelian gauge field on the D-brane as well as to
generic target space background fields. We were able to demonstrate
the possibility of concrete calculations by deriving the lowest order
system of equations of motion for both the nonabelian gauge field on the 
D-brane and the matrix valued brane position.

With this calculation we simultaneously solved the ordering problem for
all the matrix quantities involved. Of course the formalism can be extended
to higher orders, too. This requires an analysis of the renormalization of 
the system of coupled quantum fields $\xi $ and $\bar{\zeta},~\zeta $.
At the present stage it seems to be open whether the overall symmetrization 
is the correct solution of the ordering problem in higher orders, too. 
This is due to the nested structure of renormalization.

Certainly the most interesting aspect of further work concerns the search
for nontrivial dynamical effects implied by the equations of motion derived 
in this paper. We would also like to understand possible relations to recent 
work on D-branes in the context of noncommutative geometry 
\cite{banks,connes}.
\\[10mm]
{\bf Acknowledgement:}\\
I would like to thank H.-J. Otto for useful discussions. 

\end{document}